\documentclass[aps,prl,superscriptaddress,reprint]{revtex4-1}
\usepackage{amsmath}
\usepackage{amssymb}
\usepackage{graphicx}
\usepackage[colorlinks,urlcolor=blue]{hyperref}
\usepackage{url}
\usepackage{color,xcolor}
\usepackage{ulem}
\usepackage{float}
\usepackage{epstopdf}
\usepackage[none]{hyphenat}
\begin{document}

\title{Electronic structure, magnetic correlations, and superconducting pairing \\in the reduced Ruddlesden-Popper bilayer La$_3$Ni$_2$O$_6$ under pressure: \\different role of $d_{3z^2-r^2}$ orbital compared with La$_3$Ni$_2$O$_7$}
\author{Yang Zhang}
\author{Ling-Fang Lin}
\affiliation{Department of Physics and Astronomy, University of Tennessee, Knoxville, Tennessee 37996, USA}
\author{Adriana Moreo}
\affiliation{Department of Physics and Astronomy, University of Tennessee, Knoxville, Tennessee 37996, USA}
\affiliation{Materials Science and Technology Division, Oak Ridge National Laboratory, Oak Ridge, Tennessee 37831, USA}
\author{Thomas A. Maier}
\affiliation{Computational Sciences and Engineering Division, Oak Ridge National Laboratory, Oak Ridge, Tennessee 37831, USA}
\author{Elbio Dagotto}
\affiliation{Department of Physics and Astronomy, University of Tennessee, Knoxville, Tennessee 37996, USA}
\affiliation{Materials Science and Technology Division, Oak Ridge National Laboratory, Oak Ridge, Tennessee 37831, USA}
\date{\today}

\begin{abstract}
The recent discovery of superconductivity in bilayer La$_3$Ni$_2$O$_7$ (327-LNO) under pressure stimulated much interest in layered nickelates. However, superconductivity was not found in another bilayer nickelate system, La$_3$Ni$_2$O$_6$ (326-LNO), even under pressure. To understand the similarities and differences between 326-LNO and 327-LNO, using density functional theory and the random phase approximation (RPA), we systematically investigate 326-LNO under pressure. The large crystal-field splitting between the $e_g$ orbitals caused by the missing apical oxygen moves the $d_{3z^2-r^2}$ orbital farther away from the Fermi level, implying that the $d_{3z^2-r^2}$ orbital plays a less important role in 326-LNO than in 327-LNO. This also results in a smaller bandwidth for the $d_{x^2-y^2}$ orbital and a reduced energy gap for the bonding-antibonding splitting of the $d_{3z^2-r^2}$ orbital in 326-LNO, as compared to 327-LNO. Moreover, the in-plane hybridization between the $d_{x^2-y^2}$ and $d_{3z^2-r^2}$ orbitals is found to be small in 326-LNO, while it is much stronger in 327-LNO. Furthermore, the low-spin ferromagnetic (FM) state is found to be the likely ground state in 326-LNO under high pressure. The weak inter-layer coupling suggests that $s_{\pm}$-wave pairing is unlikely in 326-LNO. The robust in-plane ferromagnetic coupling also suggests that $d$-wave superconductivity, which is usually caused by antiferromagnetic fluctuations of the $d_{x^2-y^2}$ orbital, is also unlikely in 326-LNO. These conclusions are supported by our many-body RPA calculations of the pairing behavior. Contrasting with the cuprates, for the bilayer cuprate HgBa$_2$CaCu$_2$O$_6$, we find a strong ``self-doping effect'' of the $d_{x^2-y^2}$ orbital under pressure, with the charge of Cu being reduced by approximately 0.13 electrons from 0 GPa to 25 GPa. In contrast, we do not observe such a change in the electronic density in 326-LNO under pressure, establishing another important difference between the nickelates and the cuprates.
\end{abstract}

\maketitle
\section{I. Introduction}
\begin{figure*}
\centering
\includegraphics[width=0.92\textwidth]{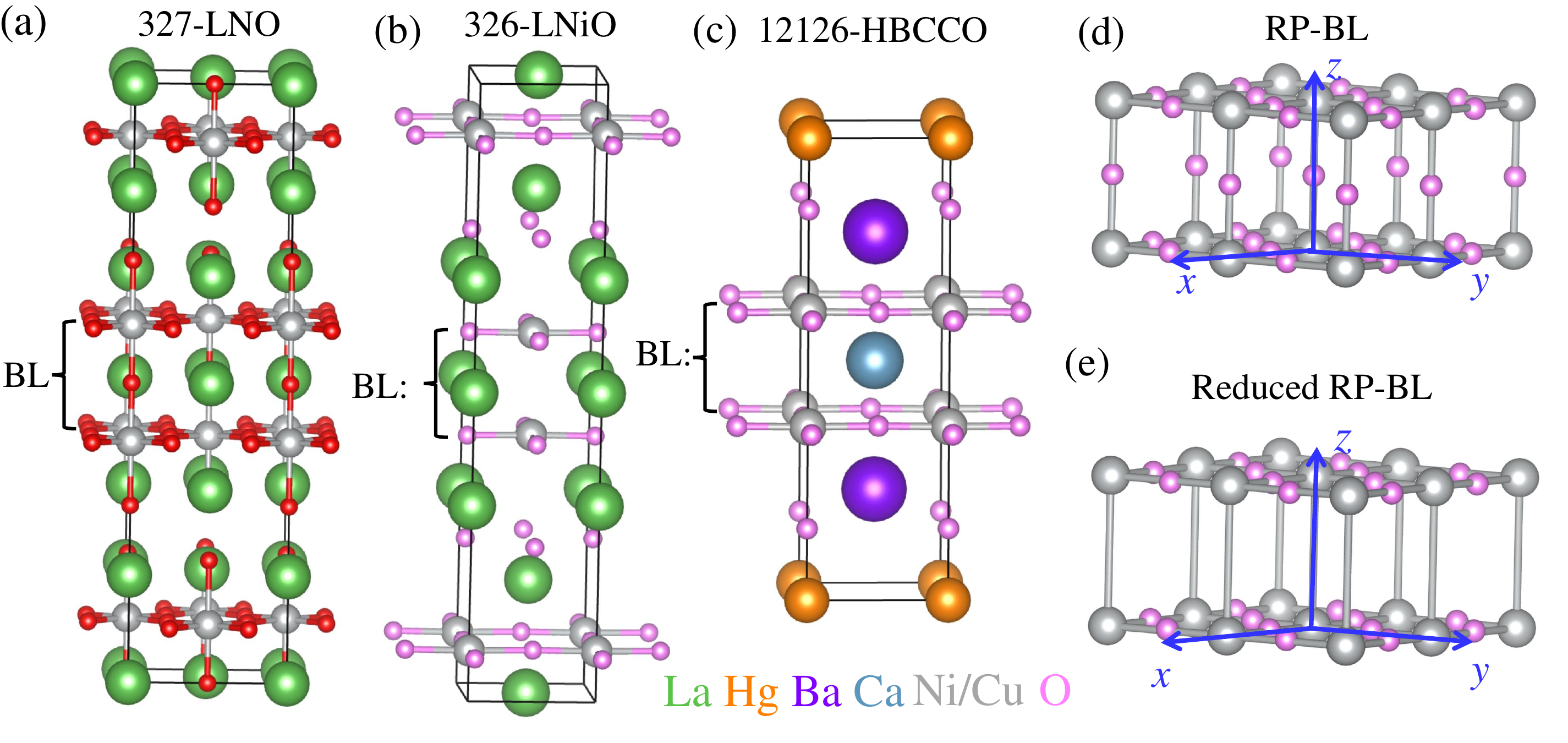}
\caption{(a-c) Schematic crystal structures of the conventional cell of (a) 327-LNO, (b) 326-LNO, and (c) 12126-HBCCO (green: La; orange: Hg;  violet: Ba; sky blue: Ca; gray: Ni or Cu; pink: O). (d-e) Sketch of the bilayer structures studied here: (d)
RP-BL sublattice with NiO$_6$ plane; (e) Reduced RP-BL sublattice with NiO$_4$ or CuO$_4$ planes. The local $z$-axis is perpendicular to the planes, while the local $x$- or $y$-axis are along the in-plane Ni-O or Cu-O bond directions, resulting in $d_{x^2-y^2}$ orbitals lying in the layer plane.}
\label{crystal}
\end{figure*}

The recent experimental discovery of pressure-induced superconductivity in the Ruddlesden-Popper bilayer (RP-BL) perovskite La$_3$Ni$_2$O$_7$ (327-LNO)~\cite{Sun:arxiv} opened a novel platform
for understanding and studying layered nickel-based high temperature superconductors~\cite{Luo:arxiv,Zhang:arxiv,Yang:arxiv,Sakakibara:arxiv,Gu:arxiv,Shen:arxiv,Liu:arxiv,Zhang:arxiv1,Lu:arxiv,Oh:arxiv,
Liao:arxiv,Qu:arxiv,Yang:arxiv1,Cao:arxiv,Lechermann:arxiv,Christiansson:arxiv,LiuZhe:arxiv,Wu:arxiv,Shilenko:arxiv,Chen:arxiv,Zhang:arxiv-exp,
Hou:arxiv,Jiang:arxiv,Huang:arxiv,Zhang:prb23,Qin:arxiv,Tian:arxiv,Lu:arxiv08,Jiang:arxiv08,Luo:arxiv08,Yang:arxiv09,Zhang:arxiv09,Zhang:arxiv9,Pan:arxiv09,Geisler:arxiv,
Yang:arxiv9,Rhodes:arxiv,LaBollita:arxiv,Wang:arxiv9,Kaneko:arxiv9,Lu:arxiv09}. The compound 327-LNO has an orthorhombic structure with a stacked bilayer NiO$_6$ octahedron sublattice geometry, where superconductivity with
the highest $T_c$ up to 80~K was reported in the high-pressure phase [see Fig.~\ref{crystal}(a)]~\cite{Sun:arxiv}.

Under the influence of hydrostatic pressure, the structure of 327-LNO transforms from the Amam to the Fmmm symmetry followed by the stabilization of a superconducting phase for a broad range of pressures from 14 to 43.5 GPa~\cite{Sun:arxiv}. The electronic density of Ni is $n = 7.5$  in 327-LNO, corresponding to Ni$^{2.5+}$ on average, resulting in two $e_g$ orbitals ($d_{x^2-y^2}$ and $d_{3z^2-r^2}$) contributing to the Fermi surface (FS) based on density functional theory (DFT) calculations~\cite{Luo:arxiv,Zhang:arxiv}. The $d_{x^2-y^2}$ orbital is nearly quarter-filled and the $d_{3z^2-r^2}$ orbital is closed to half-filled, establishing a two-orbital minimum model. In addition, the partial nesting of the FS for wavevectors ($\pi$,0) and (0,$\pi$) favors $s_{\pm}$-wave superconductivity induced by the
strong inter-layer coupling in 327-LNO, as discussed in recent theoretical efforts~\cite{Yang:arxiv,Sakakibara:arxiv,Gu:arxiv,Liu:arxiv,Zhang:arxiv1,Liao:arxiv,Qu:arxiv,Luo:arxiv08,Lu:arxiv08,Yang:arxiv9,Pan:arxiv09}. Other studies alternatively suggest the possibility of $d$-wave pairing superconductivity~\cite{Jiang:arxiv08,Geisler:arxiv,Jiang:arxiv}, as in the cuprates.

By chemical reduction, namely removal of the apical oxygen from 327-LNO, the compound
La$_3$Ni$_2$O$_6$ (326-LNO) was obtained experimentally. 326-LNO also has a stacking Ni bilayer structure [see Fig.~\ref{crystal}(b)] but displays the NiO$_2$ square-planar bilayer sublattice. The material 326-LNO reminds us of the previously well-studied bilayer superconducting cuprates HgBa$_2$CaCu$_2$O$_6$ (12126-HBCCO)~\cite{Loureiro:pc}, with a similar CuO$_2$ square-planar bilayer sublattice [see Fig.~\ref{crystal}(c)].
The difference between reduced RP-BL and  RP-BL structures is that there are no additional apical O atoms
connecting two Ni or Cu layers in the reduced RP-BL lattice, as shown in Figs.~\ref{crystal}(d) and (e).

At ambient pressure, no magnetic order was found down to 4~K for 326-LNO~\cite{Poltavets:prl09} but the existence of magnetic correlations was observed in the powder samples by nuclear magnetic resonance~\cite{apRoberts-Warren:prb}. Furthermore, weak ferromagnetic (FM) tendencies were also reported in 326-LNO at 5~K that persist at least up to 400 K~\cite{Poltavets:prl09}. This is considered to be related to nearly degenerate FM and antiferromagnetic (AFM) states~\cite{Poltavets:prl09,Sarkar:prb}. In addition, a checkerboard charge-ordered insulating state with AFM coupling was also predicted in 326-LNO~\cite{Botana:prb}. However, due to the intrinsic limitations of powder samples~\cite{Poltavets:prl09}, the AFM charge-order instability was not confirmed yet. Of primary importance for the work discussed here, contrary to the pressure-induced superconductivity of 327-LNO, superconductivity was not observed in 326-LNO under pressure up to a
maximum of 25.3 GPa, although an insulator-metal transition was found around 6.1 GPa in recent experiments~\cite{Liu:scpma}.

Considering these studies in bilayer systems, several interesting questions naturally arise: what are the similarities and differences between the bilayer 326-LNO and 327-LNO nickelates under pressure? What causes these differences? Does the missing apical oxygen in 326-LNO play a key role in the reported absence of superconductivity under pressure?
What is the connection between 326-LNO and 12126-HBCCO?

To address these questions, here we theoretically studied the 326-LNO compound under pressure, by using first-principles DFT as well as random phase approximation (RPA) calculations. Similarly to 327-LNO, pressure increases the bandwidth of the Ni's $3d$ states, leading to an enhanced itinerant behavior and thus effectively reduced electronic correlations. Furthermore, the Ni's $3d$ orbitals are mainly located near the Fermi level and most of the O's $2p$ states are far away from that Fermi level, indicating a robust charge-transfer energy ($\varepsilon_d$ - $\varepsilon_O$) in both 326-LNO and 327-LNO, establishing a common character among these nickelates.  In addition, the $d_{3z^2-r^2}$ orbital displays a bonding-antibonding splitting character in both 326-LNO and 327-LNO, as well as in 12126-HBCCO.

However, different from 327-LNO, the crystal-field splitting between the $e_g$ orbitals is much larger in 326-LNO
and also the in-plane hybridization between the $d_{x^2-y^2}$ and $d_{3z^2-r^2}$ orbitals was found to be very small in the latter, leading to only two Fermi surface sheets, $\alpha$ and $\beta$, composed primarily of the single $d_{x^2-y^2}$ orbital.  By introducing electronic correlations, the low-spin FM state was found to have the lowest energy among the five considered candidates under pressure, with a very weak magnetic coupling between the layers. This strongly suggests that $s_{\pm}$-wave pairing is unlikely in 326-LNO. Furthermore, the large in-plane FM coupling also indicates that $d$-wave superconductivity, usually caused by AFM fluctuations of the $d_{x^2-y^2}$ orbital, is also unlikely in 326-LNO. These qualitative conclusions are supported by our many-body RPA calculations. In addition, we do not observe any obvious changes of the electronic density in 326-LNO under pressure, while there is a strong  ``self-doping effect'' of the $d_{x^2-y^2}$ orbital in 12126-HBCCO, establishing another difference between the nickelates and the cuprates.

\section{II. Method}
In the present study, the first-principles DFT calculations were performed by using the Vienna {\it ab initio} simulation package (VASP) code, within the projector augmented wave (PAW) method~\cite{Kresse:Prb,Kresse:Prb96,Blochl:Prb}, with the generalized gradient approximation and the Perdew-Burke-Ernzerhof (PBE) exchange potential~\cite{Perdew:Prl}. The plane-wave cutoff energy was set as $550$~eV.

Both lattice constants and atomic positions were fully relaxed until the Hellman-Feynman force on each atom was smaller than $0.01$ eV/{\AA}. The $k$-point mesh was appropriately modified for different crystal structures to make the $k$-point densities approximately the same in reciprocal space (e.g. $16\times16\times3$ for the conventional structure of 326-LNO in the I4/mmm phase). In addition to the standard DFT calculation, we employed the maximally localized Wannier functions (MLWFs) method to fit the Ni's $e_g$ bands to obtain the hoppings and crystal-field splittings for our subsequent model RPA calculations, as well as obtaining the FSs, using the WANNIER90 packages~\cite{Mostofi:cpc}. Furthermore, the calculated three-dimensional FSs obtained from MLWFs were visualization
by XCRYSDEN package~\cite{XCrySDen}. All crystal structures were visualized with the VESTA code~\cite{Momma:vesta}.

To discuss the magnetic tendencies in 327-LNO under pressure, a strong intra-atomic interaction was considered in a screened Hartree-Fock-like manner, as used in the local density approach (LDA) plus $U$ method with Liechtenstein format within the double-counting item~\cite{Liechtenstein:prb}. In addition, specific values for $U = 4.75$ eV and $J = 0.68$ eV were considered in our study of 326-LNO, as used in a previous study~\cite{Botana:prb,Pardo:prb}.

To investigate the superconducting pairing properties of the 326-LNO system, we first constructed a four-band $e_g$ orbital tight-binding model on a bilayer lattice~\cite{Maier:prb11,Mishra:sr,Maier:prb19,Maier:prb22}, involving two Ni sites with $e_g$ orbitals in a unit cell with an overall filling of $n = 5$.
The kinetic hopping component of the Hamiltonian is
\begin{eqnarray} \label{eq:Htb}
H_k = \sum_{\substack{i\sigma\\\vec{\alpha}\gamma\gamma'}}t_{\gamma\gamma'}^{\vec{\alpha}}
(c^{\dagger}_{i\sigma\gamma}c^{\phantom\dagger}_{i+\vec{\alpha}\sigma\gamma'}+H.c.)+ \sum_{i\gamma\sigma} \Delta_{\gamma} n_{i\gamma\sigma}.
\end{eqnarray}
The first term represents the hopping of an electron from orbital $\gamma$ at site $i$ to orbital $\gamma'$ at the neighboring site $i+\vec{\alpha}$. $c^{\dagger}_{i\sigma\gamma}$($c^{\phantom\dagger}_{i\sigma\gamma}$) is the standard creation (annihilation) operator, $\gamma$ and $\gamma'$ represent the different orbitals, and $\sigma$ is the $z$-axis spin projection. $\Delta_{\gamma}$ represents the crystal-field splitting of each orbital $\gamma$. The vectors $\vec{\alpha}$ are along the three bilayer-lattice directions [see Fig.~\ref{crystal}(e)], defining different neighbors of hoppings (the detailed hoppings can be found in the supplemental materials~\cite{SM}).

This Hamiltonian is supplemented with an interaction term that contains on-site intra-orbital $U$ and inter-orbital $U'$ Coulomb repulsions as well as Hund's coupling $J$ and pair-hopping $J'$ terms. To assess this model for its pairing behavior, we performed many-body RPA calculations, which are based on a perturbative weak-coupling expansion in the Coulomb interaction~\cite{Kubo2007,Graser2009,Altmeyer2016,Romer2020}. In our multi-orbital RPA technique~\cite{Kubo2007,Graser2009,Altmeyer2016}, the RPA enhanced spin susceptibility is obtained from the Lindhart function $\chi_0({\bf q})$:
\begin{eqnarray}
\chi({\bf q}) = \chi_0({\bf q})[1-{\cal U}\chi_0({\bf q})]^{-1}.
\end{eqnarray}
Here, $\chi_0({\bf q})$ is an orbital-dependent susceptibility tensor and ${\cal U}$ is a tensor that contains the interaction parameters \cite{Graser2009}. The pairing strength $\lambda_\alpha$ for channel $\alpha$ and the corresponding gap structure $g_\alpha({\bf k})$ are obtained from solving an eigenvalue problem of the form
\begin{eqnarray}\label{eq:pp}
	\int_{FS} d{\bf k'} \, \Gamma({\bf k -  k'}) g_\alpha({\bf k'}) = \lambda_\alpha g_\alpha({\bf k})\,,
\end{eqnarray}
where the momenta ${\bf k}$ and ${\bf k'}$ are on the Fermi surface, and $\Gamma({\bf k - k'})$ contains the irreducible particle-particle vertex. In the RPA approximation, the dominant term entering $\Gamma({\bf k-k'})$ is the RPA spin susceptibility $\chi({\bf k-k'})$.

\section{III. Results}

\subsection{A. Electronic structures and Fermi surface}
In the pressure range that we studied, the electronic structures of 326-LNO remain very similar in shape,
while pressure increases the bandwidth of the Ni's $3d$ states, leading to an enhanced itinerant behavior.
This larger bandwith ``effectively'' reduces the electronic correlations (see Appendix). Here, unless otherwise specified, we will mainly focus on the results at 25 GPa to understand the similarities and differences between the bilayer 326-LNO and 327-LNO systems. At this pressure, 327-LNO is already superconducting but 326-LNO is not~\cite{FS}.

\begin{figure}
\centering
\includegraphics[width=0.46\textwidth]{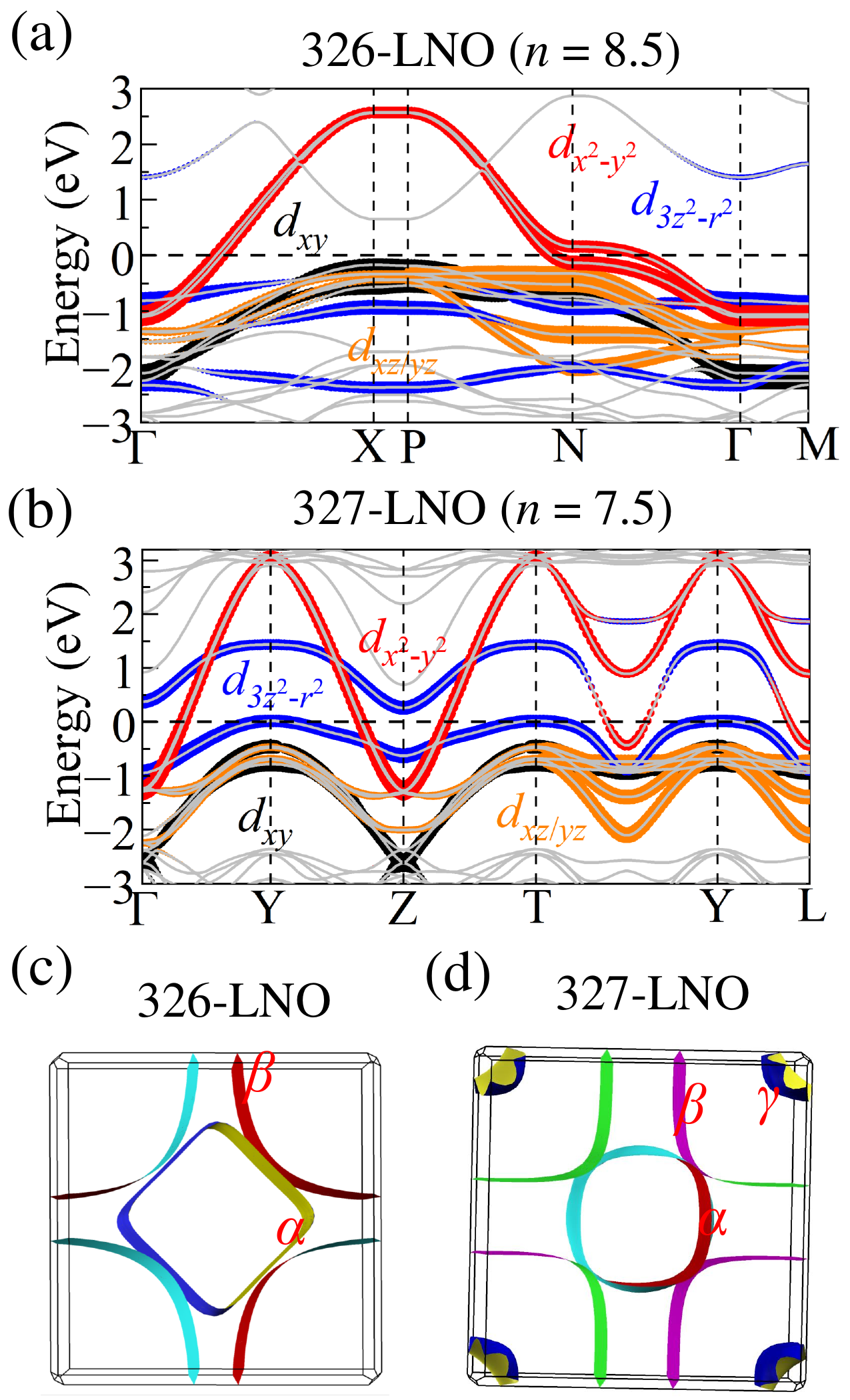}
\caption{ Projected band structures of the non-magnetic phase of (a) I4/mmmm 326-LNO and (b) Fmmm 327-LNO structures without any electronic interaction, at 25 GPa. The weight of each Ni orbital is given by the size of the circles. The Fermi level (zero energy) is marked by the horizontal dashed line. The coordinates of the high-symmetry points of the Brillouin zone are (a) $\Gamma$ = (0, 0, 0), X = (0, 0, 0.5), P = (0.25, 0.25, 0.25), N = (0, 0.5, 0), and R = (0.5, 0.5, -0.5) for I4/mmmm 326-LNO and (b) $\Gamma$ = (0, 0, 0), Y = (0.5, 0, 0.5), Z = (0.5, 0.5, 0), T = (0, 0.5, 0.5), and L = (0.5, 0.5, 0.5) for Fmmm 327-LNO. FSs of the (c) I4/mmmm 326-LNO and (d) Fmmm 327-LNO at 25 GPa. The existence of
$\gamma$ pockets is clearly visible in 327-LNO but they are absent in 326-LNO.}
\label{Pband-fs}
\end{figure}

As shown in Figs.~\ref{Pband-fs} (a) and (b), in both 326-LNO and 327-LNO, the Ni's $3d$ orbitals are those that primarily contribute to the electronic density near the Fermi level, hybridized with O $p$-states. Furthermore, the O $p$-states are mainly located in lower energy regions than the Ni's $3d$ states, indicating a large charge-transfer gap between Ni's $3d$ and O's $2p$ orbitals ($\varepsilon_d$ - $\varepsilon_O$). This is similar to what we found in our previous study of the infinite-layer NdNiO$_2$~\cite{Zhang:prb20}. In addition, the three $t_{2g}$ orbitals are fully occupied, while the $d_{x^2-y^2}$ is partially occupied crossing the Fermi level in both cases of 326-LNO and 327-LNO. Compared with 327-LNO, the $d_{x^2-y^2}$ orbital is less itinerant with a reduced bandwidth of $\sim 20\%$ in the 326-LNO case, resulting in a reduced nearest-neighbor hopping for the $d_{x^2-y^2}$ orbital. This suggests that the additional apical oxygen connected to two Ni layers in 327-LNO enhances the itinerant behavior of the $d$ orbitals, reducing the ``effective'' electronic correlations $U/W$ in 327-LNO, as compared to those of 326-LNO.

Furthermore, the Ni $d_{3z^2-r^2}$ orbital shows a bonding-antibonding molecular-orbital splitting character in both cases, caused by the dimer structure in the bilayers, as discussed in 327-LNO~\cite{Zhang:arxiv}. Compared with 327-LNO, the energy gap between bonding and antibonding states decreases by about $21 \%$ in 326-LNO, indicating that the ``bridge'' of the apical oxygen would increase the hopping and enhance the bonding-antibonding splitting. Hence, in both 326-LNO and 327-LNO, the $d_{3z^2-r^2}$ states are more localized and $d_{x^2-y^2}$ states are more itinerant.

Because there are no apical oxygens connecting two Ni sites between the two layers of bilayer 326-LNO, the crystal-field splitting $\Delta$ between the orbitals $d_{3z^2-r^2}$ and $d_{x^2-y^2}$ increases singificantly ($\sim 1.96$ eV) as compared with that in 327-LNO ($\sim 0.51$ eV)~\cite{Zhang:arxiv1}. In this case, the interlayer magnetic coupling should be quite small in 326-LNO, suggesting a different role of the $d_{3z^2-r^2}$ orbital in those two systems, although they both have a bilayer Ni sublattice. Moreover, the in-plane interorbital hopping between the $e_g$ orbitals is also rather small in 326-LNO ($\sim 0.013$ eV), leading to a reduced in-plane hybridization between the $d_{3z^2-r^2}$ and $d_{x^2-y^2}$ orbitals compared to that in 327-LNO ($\sim 0.243$ eV)~\cite{Zhang:arxiv1}.

As a result of these differences, in 326-LNO, only the $d_{x^2-y^2}$ orbital contributes to the FS, leading to two strongly two-dimensional sheets ($\alpha$ and $\beta$), as shown in Fig.~\ref{Pband-fs}(c). However, the FS of 327-LNO in the Fmmm phase is made up of both $d_{x^2-y^2}$ and $d_{3z^2-r^2}$ orbitals, resulting in two sheets ($\alpha$ and $\beta$) and an additional pocket ($\gamma$). Due to the strong hybridization of the $e_g$ states in 327-LNO, the two sheets $\alpha$ and $\beta$ display a mixed character between the $d_{3z^2-r^2}$ and $d_{x^2-y^2}$ orbitals. Hence, $d_{3z^2-r^2}$ is not as important in the 326-LNO case as in 327-LNO, which may be crucial to understand the absence of superconductivity in the former.

\subsection{B. Magnetic correlations in 326-LNO under pressure}

Next, we introduced local Hubbard couplings and studied the magnetic correlations in 326-LNO under pressure. For our studies, several magnetic structures of the Ni bilayer spins were considered: (1) A-AFM: FM coupling in the NiO$_2$ layer plane and AFM coupling between the Ni layers; (2) FM: FM coupling along both the NiO$_2$ layer plane and between the Ni layers; (3) G-AFM: AFM coupling along both the NiO$_2$ layer plane and between the Ni layers; (4) C-AFM: AFM coupling along the NiO$_6$ layer plane and FM coupling between the layers; (5) Stripe-AFM: AFM in one in-plane direction and FM in the other, while the coupling along the Ni layers direction is AFM. For all these states we used the specific values $U = 4.75$ eV and $J = 0.68$ eV for 326-LNO in the LDA+$U$ format with double-counting item~\cite{Liechtenstein:prb}, as used in a previous study of 326-LNO~\cite{Botana:prb,Pardo:prb}.

\begin{figure}
\centering
\includegraphics[width=0.46\textwidth]{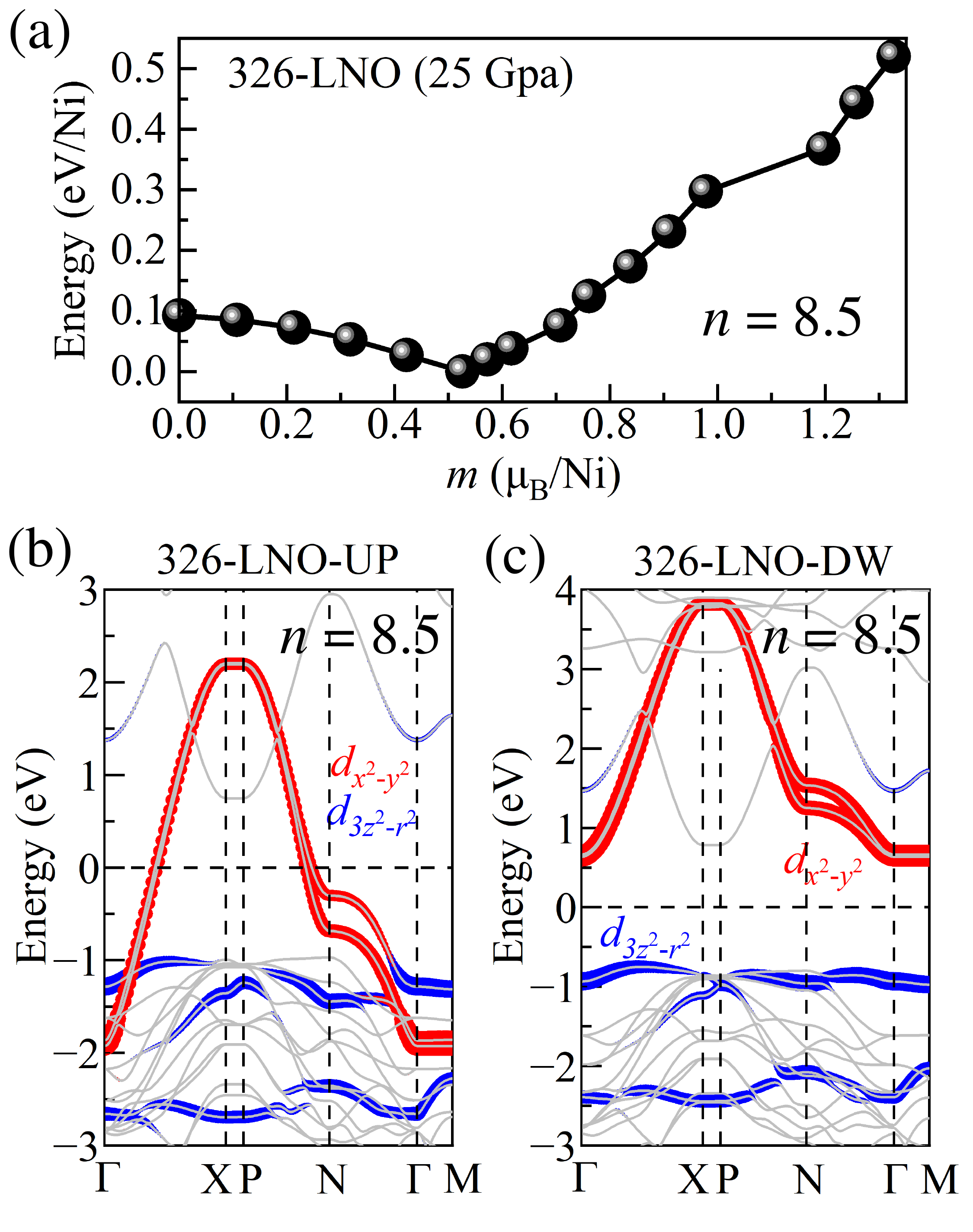}
\caption{(a) Total energy as a function of the magnetic moment of the Ni ions for 326-LNO at 25 GPa. (b-c) Projected band structures of the FM phase in the I4/mmmm structural phase of 326-LNO for (b) spin-up and (c) spin-down at 25 GPa. The key $e_g$ orbitals are marked by the red and blue colors. The Fermi level (zero energy) is marked by the horizontal dashed line.}
\label{magnetism}
\end{figure}

\begin{table}
\caption{Energy differences (meV/Ni) and calculated magnetic moment ($\mu$$_B$/Ni) for the various input spin configurations used here. The FM configuration is taken as the energy of reference.}
\begin{tabular*}{0.48\textwidth}{@{\extracolsep{\fill}}lllc}
\hline
\hline
Magnetism & Energy & Magnetic moment  \\
\hline
A-AFM     & 0.65 & 0.530   \\
FM        & 0  & 0.526   \\
G-AFM & 107.92  & 0.208  \\
C-AFM & 104.00  & 0.212  \\
Stripe     & 43.91  & 0.455    \\
\hline
\hline
\end{tabular*}
\label{Table1}
\end{table}

Considering the $d^{8.5}$ electronic configuration in 326-LNO and the square-planar crystal-field splitting, the Ni ions are expected to be in a low-spin state for 326-LNO. To confirm this, we calculated the total energy as a function of the magnetic moment of the Ni ions for 326-LNO at 25 GPa, using the fixed-spin-moment method. Figure~\ref{magnetism}(a) clearly shows an energy minimum around 0.53 $\mu$$_B$/Ni, supporting the low-spin picture in 326-LNO.

Next, using the same crystal structure, we calculated the energies for different magnetic configurations. As shown in Table~\ref{Table1}, the FM state has the lowest energy among the five considered candidates. In addition, the energy difference between the A-AFM and FM states is quite small, indicating that the coupling between layers is weak in 326-LNO, while a strong AFM coupling was found in 327-LNO due to the large hopping amplitude of the $d_{3z^2-r^2}$ orbital between the layers~\cite{Zhang:arxiv,Zhang:arxiv1}.  The weak inter-layer FM coupling in 326-LNO suggests that $s_{\pm}$-wave pairing discussed in the context of 327-LNO may not be favored. Moreover, the C-AFM state has a much higher energy than the FM phase, indicating a large in-plane FM coupling in 326-LNO, while the in-plane magnetic coupling is much weaker in 327-LNO~\cite{Zhang:arxiv,Zhang:arxiv1}. The in-plane strong FM coupling is expected to disfavor $d$-wave superconductivity, which is induced by in-plane AFM fluctuations of the $d_{x^2-y^2}$ orbitals. These considerations suggest that 326-LNO is far from a superconductiing instability.

Furthermore, we also calculated the band structure of the FM state for 326-LNO at 25 GPa. As shown in Figs.~\ref{magnetism}(b) and (c), the fully-occupied $d_{3z^2-r^2}$ states have Mott-localized characteristics in both the bonding and antibonding states in 326-LNO far from the Fermi level. However, the spin-up and spin-down states are well-separated for the Ni's $d_{x^2-y^2}$ but with a fractional occupation of the spin-up bands, leading to metallic behavior.

\subsection{C. RPA results for 326-LNO}

As discussed in the previous section, based on calculations of magnetism, both the $s_{\pm}$-wave and $d$-wave pairing appear to be unlikely in 326-LNO. To better understand the superconducting pairing in 326-LNO, we performed multi-orbital RPA calculations for the four-band $e_g$ bilayer tight-binding model in Eq.~(\ref{eq:Htb}).

\begin{figure}
\centering
\includegraphics[width=0.45\textwidth]{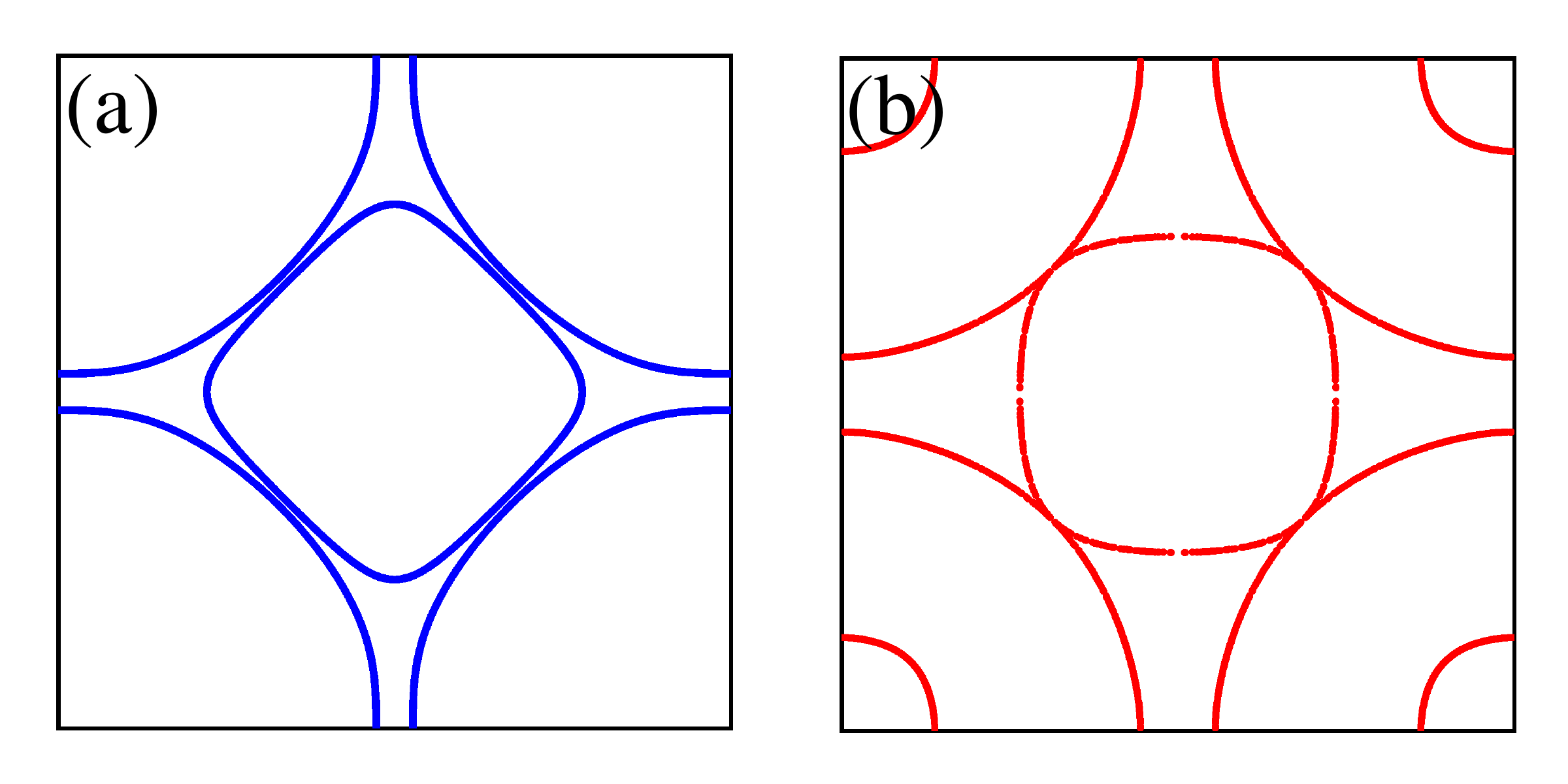}
\includegraphics[width=0.50\textwidth]{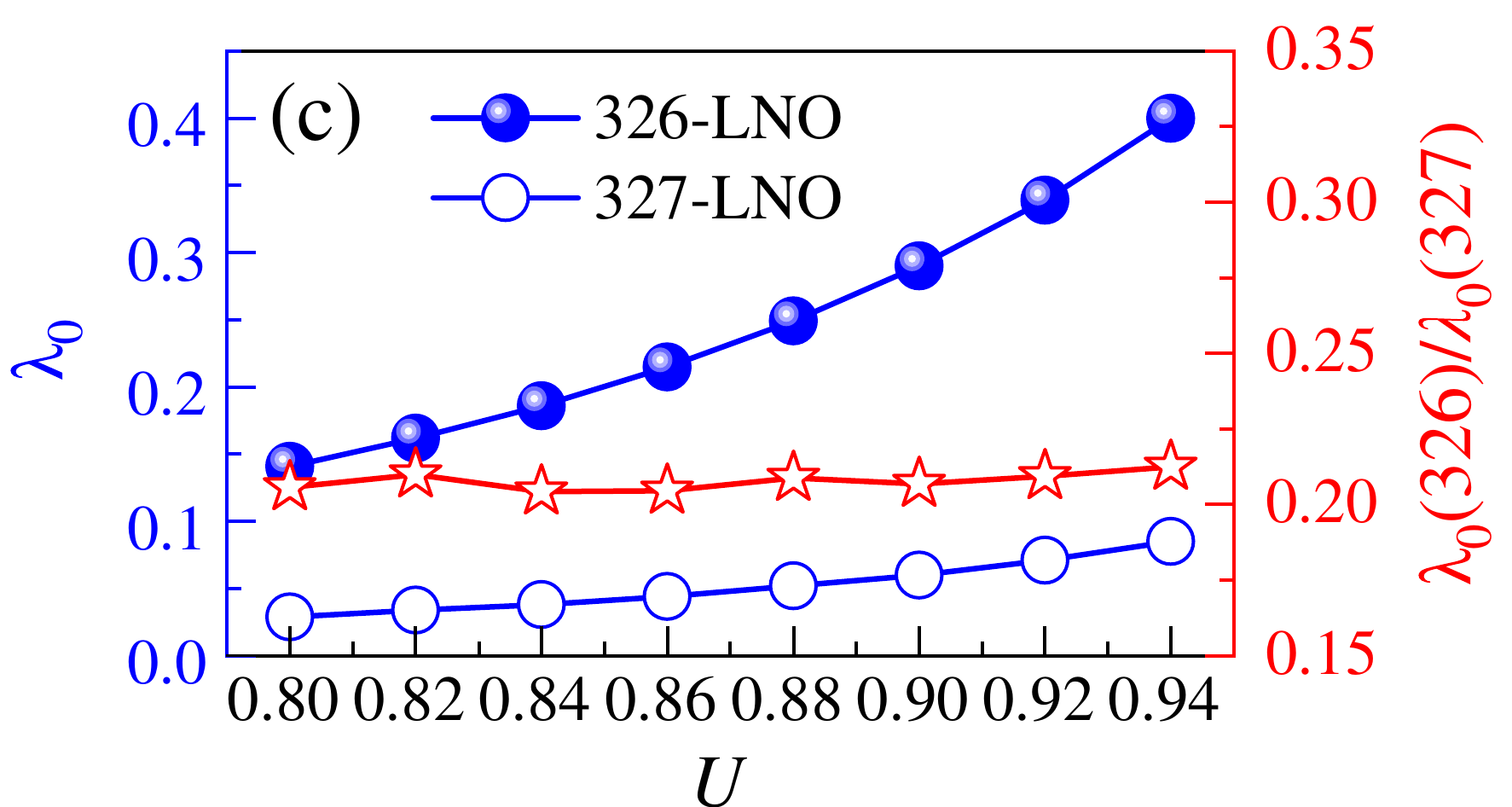}
\caption{(a-b) FSs obtained from tight-binding calculations. (a) The hopping and crystal field were obtained from 326-LNO at 25 GPa. Here, the four-band $e_g$ orbital model was considered with the overall filling $n = 5$ (2.5 electrons per site) in a bilayer lattice. Furthermore, we considered three neighbors of hopping in the $xy$ plane and two neighbors of hopping between layers in our model calculation, while other long-range hoppings are ignored. (b) The hopping and crystal field were obtained for 327-LNO at 25 GPa~\cite{Zhang:arxiv1}. (c) The RPA calculated pairing strength $\lambda_0$ of the leading $g$-wave pairing state found for 326-LNO and the $s_{\pm}$-wave state for 327-LNO as a function of $U$ (in units of eV) obtained from the bilayer model. Here we have set $U'=U/2$ and $J=J'=U/4$. The ratio $\lambda_0$(326-LNO)/$\lambda_0$(327-LNO) is also presented in red.}
\label{rpa}
\end{figure}

As shown in Fig.~\ref{rpa}(a), the FS obtained from the tight-binding model fits the DFT FS well, consisting of two sheets ($\alpha$ and $\beta$) made up primarily of the $d_{x^2-y^2}$ orbital. As a comparison, the model FS of 327-LNO is also shown in Fig.~\ref{rpa}(b), where the hoppings, overall filling, and crystal field splitting were taken from our previous study~\cite{Zhang:arxiv1}.

Figure~\ref{rpa}(c) shows the RPA results for the pairing strength $\lambda_0$ of the leading pairing instability calculated from Eq.~(\ref{eq:pp}) for 326-LNO and 327-LNO as a function of the intra-orbital Coulomb repulsion $U$. Here we have set the inter-orbital Coulomb repulsion $U'=U/2$ and the Hund's rule coupling and pair hopping $J=J'=U/4$. For 327-LNO, the leading pairing state has $s_\pm$ symmetry, as we discussed before in Ref.~\cite{Zhang:arxiv1}, for all values of $U$. For 326-LNO, both $s_\pm$ and $d_{x^2-y^2}$-wave states are not competitive, and we instead find a leading $g$-wave state (not shown) for all values of $U$. As expected, in both cases, $\lambda_0$ increases with increasing $U$. More importantly, however, the leading $g$-wave state for 326-LNO has a significantly lower $\lambda_0$ (by about a factor of 5) compared to that of the leading $s_{\pm}$ state for 327-LNO. This provides evidence for a substantial qualitative difference between the two systems and shows that the 326-LNO system is far from a superconducting instability.

As discussed before, one can understand the absence of an $s_\pm$ instability in 326-LNO from the fact that, compared to 327-LNO, the $d_{3z^2-r^2}$ orbital is much farther from the Fermi level, and therefore does not contribute to the low-energy physics, resulting in a much weaker inter-layer coupling. In addition, the $d_{x^2-y^2}$ orbital is at 1/4 filling in 326-LNO. This electronic density is far from the typical density region for which a single-band system like the cuprates displays $d$-wave superconductivity.

\subsection{D. Electronic structure of reduced bilayer 12126-HBCCO}

All the above discussions of 326-LNO suggest that the reduced 326 RP-BL system is very different from the 327 RP-BL system. The main
reason for this difference is that the $d_{3z^2-r^2}$ orbital plays a different role in these two bilayer systems due to different crystal-field splittings $\Delta$ of the $e_g$ orbitals with or without apical oxygen atoms. Next, let us briefly reexamine the typical reduced bilayer cuprate 12126-HBCCO to better understand the similarities and differences between the bilayer nickelates and cuprates.

\begin{figure}
\centering
\includegraphics[width=0.46\textwidth]{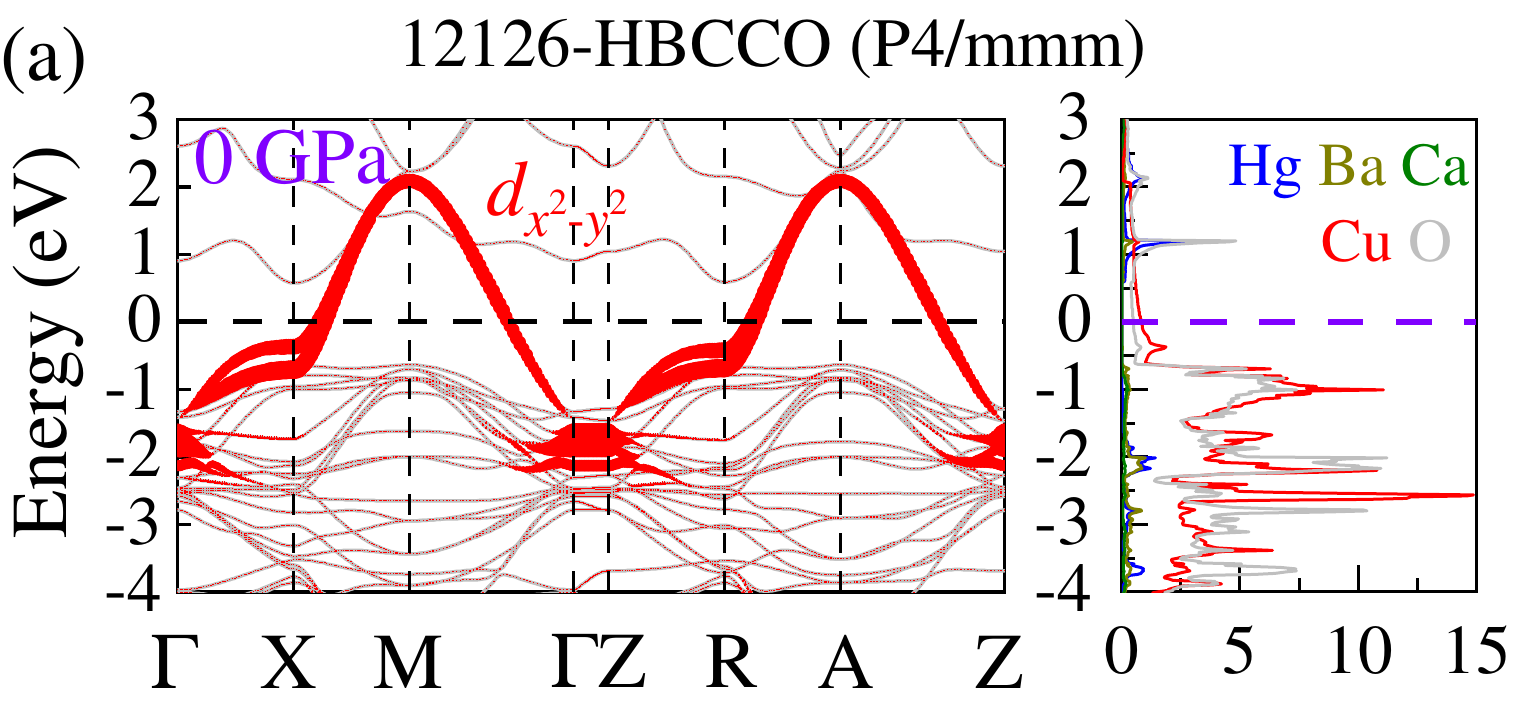}
\includegraphics[width=0.46\textwidth]{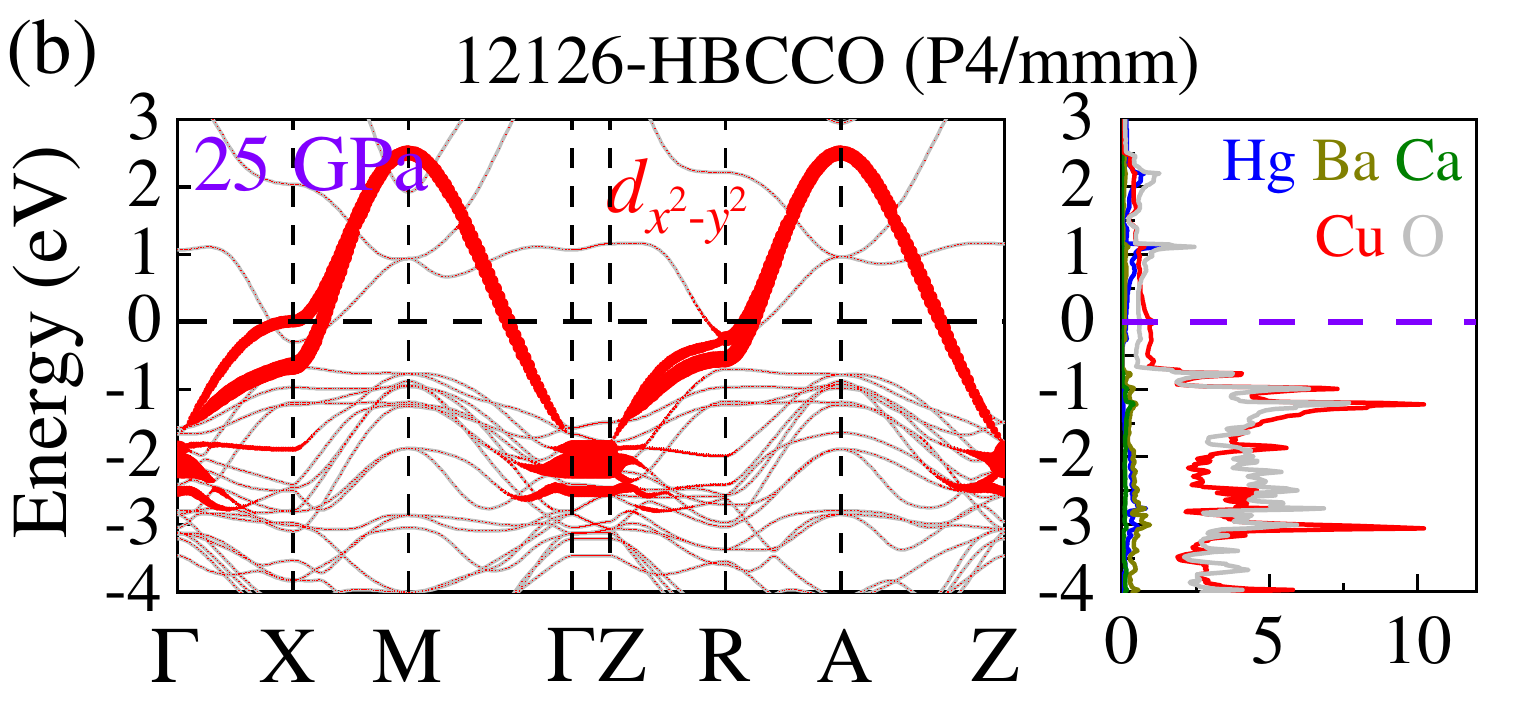}
\caption{Projected band structures and density of states for the non-magnetic state without interaction of 12126-HBCCO at (a) 0 GPa, and (b) 25 GPa, respectively. The Fermi level (zero energy) is marked by the horizontal dashed line. The coordinates of the high-symmetry points of the Brillouin zone of P4/mmmm 12126-HBCCO are $\Gamma$ = (0, 0, 0), X = (0.5, 0, 0), M = (0.5, 0.5, 0.5), Z = (0, 0, 0.5), R = (0, 0.5, 0.5), and A = (0.5, 0.5, 0.5).}
\label{Cu}
\end{figure}

12126-HBCCO forms a $P4/mmm$ tetragonal crystal structure with space group No. 123~\cite{Loureiro:pc}, as shown in Fig.~\ref{crystal}(c). Without pressure, near the Fermi level, the Bloch states are mainly composed of the Cu $3d$ orbitals that have much stronger hybridization with the O $2p$ orbitals than in 326-LNO and 327-LNO, as displayed in Fig.~\ref{Cu}(a). Furthermore, this also suggests a smaller charge-transfer gap between the Cu $d$ and $O$ $p$-states than that in 327-LNO and 326-LNO, establishing the most fundamental universal difference between nickelates and cuprates. In addition, for 12126-HBCCO, the Fermi states have contributions from the Cu $d_{x^2-y^2}$ orbital. For both RP-BL and reduced RP-BL systems, the $d_{3z^2-r^2}$ orbital has a large hopping between the two Ni layers, while the $d_{x^2-y^2}$ has zero hopping because it lies in the $xy$ plane. Hence, the Cu $d_{3z^2-r^2}$ orbital also displays a bonding-antibonding molecular-orbital splitting behavior with fully-occupied character due to a large crystal-field splitting $\Delta$ of $e_g$ orbitals, similar to 326-LNO, resulting in significant differences between RP-BL and reduced RP-BL systems. In this case, the bonding-antibonding state arises from the overlap between the $d_{3z^2-r^2}$ orbitals due to the bilayer geometry, where the apical O $p_z$ is a ``bridge'' connecting two Ni sites in the 327-LNO that can enhance the bonding-antibonding splitting.

As displayed in Fig.~\ref{Cu}(b), with increasing pressure, the bandwidth of the $d_{x^2-y^2}$ orbital of 12126-HBCCO substantially increases as compared to that at 0 GPa, implying an enhancement of the itinerant properties of the $3d$ electrons. Furthermore, the band structure of 12126-HBCCO also clearly indicates a ``self-hole-doping'' effect of the $d_{x^2-y^2}$ orbitals under pressure. According to the Bader charge analysis~\cite{Bader,Tang:jpcm,Henkelman:cms}, the charge of Cu significantly decreases by about 0.13 electrons from 0 GPa to 25 GPa. This pressure induced change of the electronic density is reminiscent of the previously studied two-leg iron ladder superconductors BaFe$_2$$X$$_3$ ($X$ = S or Se)~\cite{Zhang:prb17,Zhang:prb18}, where the self-doping effect under pressure induces superconductivity, as discussed previously using a two-orbital Hubbard model~\cite{Patel:prb16,Patel:prb16}. However, in our study we do not find any obvious significant charge transfer for 326-LNO under pressure, establishing another important difference between the nickelates and the cuprates.

\section{IV. Conclusions}

In summary, here we have systematically studied the similarities and differences of the two bilayer nickelates 326-LNO and 327-LNO. We presented our rationale for the absence of superconductivity in 326-LNO under pressure, by using DFT and RPA calculations. For both bilayer nickelates, the states near the Fermi level mainly arise from the Ni $3d$ orbitals, while most of the O $2p$ states are localized away from the Fermi energy. In addition, pressure increases the bandwidth of the Ni $3d$ states, leading to an enhanced itinerant behavior, which produces a reduced ``effective'' electronic correlation. In both 326-LNO and 327-LNO, the $d_{3z^2-r^2}$ orbital shows bonding-antibonding splitting states. The absence of the apical oxygen leads to a large crystal-field splitting between the $e_g$ orbitals in 326-LNO, resulting in the $d_{3z^2-r^2}$ orbital being far away from the Fermi level and thus reducing its importance. This also results in a smaller bandwidth for the $d_{x^2-y^2}$ orbital and a reduced bonding-antibonding energy splitting of the $d_{3z^2-r^2}$ orbital, as compared to 327-LNO. Moreover, the in-plane hybridization between $d_{x^2-y^2}$ and $d_{3z^2-r^2}$ is found to be very small in 326-LNO, much smaller than in 327-LNO.

In addition, using RPA calculations, we have found that superconducting pairing correlations are significantly weaker in 326-LNO relative to 327-LNO. Due to a much reduced inter-layer coupling, the leading $s_\pm$-wave state found for 327-LNO is suppressed for 326-LNO. Moreover, we have found that a low-spin FM state has the lowest energy among the five magnetic configurations studied, much lower than the C-AFM state, indicating a large in-plane FM coupling in 326-LNO. This, and the fact that the in-plane $d_{x^2-y^2}$ orbitals are quarter-filled, explains why AFM fluctuation driven $d$-wave pairing correlations are similarly suppressed in 326-LNO.

Similar to 326-LNO, the $d_{3z^2-r^2}$ orbital also displays a bonding-antibonding state splitting character in the cuprate 12126-HBCCO, suggesting a common electronic structure in the bilayer lattice. However, as a fundamental difference between cuprates and nickelates, the Cu $3d$ orbitals in the cuprates are highly hybridized with the O $2p$ orbitals, leading to a much smaller charge-transfer gap. Moreover, we found a strong  pressure induced ``self-doping effect'' of the $d_{x^2-y^2}$ orbital in 12126-HBCCO, where the charge of the Cu states is significantly reduced by about 0.13 electrons when changing pressure from 0 GPa to 25 GPa. However, we do not observe such a change of the electronic density in 326-LNO under pressure, indicating another important difference between the nickelate and the cuprate bilayer systems.

\section{V. Acknowledgments}
This work was supported by the U.S. Department of Energy, Office of Science, Basic Energy Sciences, Materials Sciences and Engineering Division.

\section{VI. APPENDIX}
\begin{figure}
\centering
\includegraphics[width=0.46\textwidth]{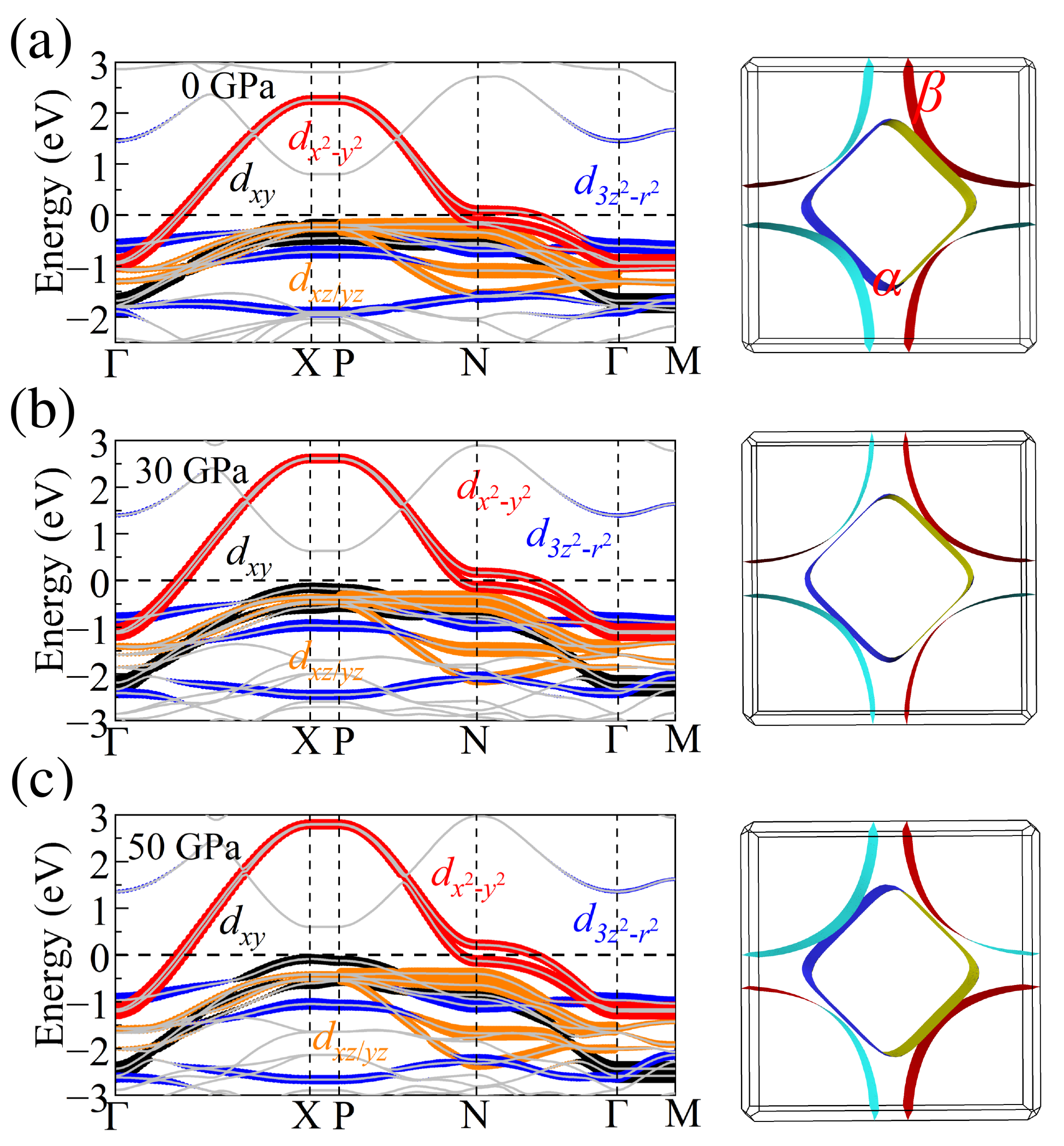}
\caption{Projected band structures and FSs of the non-magnetic phase of the I4/mmmm 326-LNO structures without any interaction at (a) 0 GPa, (b) 30 GPa, and (c) 50 GPa, respectively. The weight of each Ni orbital is given by the size of the circles. The Fermi level (zero energy) is marked by the horizontal dashed line. The coordinates of the high-symmetry points of the Brillouin zone are $\Gamma$ = (0, 0, 0), X = (0, 0, 0.5), P = (0.25, 0.25, 0.25), N = (0, 0.5, 0), and R = (0.5, 0.5, -0.5).}
\label{326-LNO}
\end{figure}

As shown in Fig.~\ref{326-LNO}, the electronic structures of 326-LNO are very similar under pressure. The $d_{x^2-y^2}$ orbital contributes the most to the Fermi level, while other $d$ orbitals are fully occupied. The FS are contributed by two sheets ($\alpha$ and $\beta$) made almost entirely of the single $d_{x^2-y^2}$ orbital.  The pressure increases the bandwidth of Ni's $3d$ states, leading to an enhanced itinerant behavior, thus also leading to reduced ``effective'' electronic correlations $U/W$.

\begin{references}
\bibitem{FS}{Here, the primitive unit cells with two Ni atoms were considered to calculate band structures and FSs. For 326-LNO, the lattice constants and atomic positions were fully relaxed within the PAW method with the PBE exchange potential. For 327-type, we used the optimized structure obtained from our previous work~\cite{Zhang:arxiv1}.}
\end{references}
\end{document}